\documentclass[aps,showpacs,twocolumn]{revtex4}
\usepackage{graphics}
\begin{document}
\newcommand{\bstfile}{aps} %alternative styles: osa, prasty or revtex
\newcommand{\bibs}{c:/PCTeXv4/TexBiB/final}
\draft
\title{Fundamental Limits of Electronic Nonlinear-Optical Phenomena}
\author{Mark G. Kuzyk}
\address{Department of Physics and Astronomy, Washington State University \\ Pullman,
Washington  99164-2814}
\date{\today}

\begin{abstract}
Using sum rules and a new dipole-free sum-over-states expression, we introduce a method for calculating the fundamental limits of the dispersion of the real and imaginary parts of all electronic nonlinear-optical susceptibilities.  As such, these general results can be used to study any nonlinear optical phenomena at any wavelength, making it possible to push both applications and our understanding of such processes to the limits.  These results reveal the ultimate constraints imposed by nature on our ability to control and use light.
 
\end{abstract}


\maketitle

%OCIS: 190.0190, 020.0020, 020.4900

\vspace{1em}

\section{Introduction}

The interaction of light with matter is of fundamental importance in studying materials;\cite{slepk04.01} and, lies at the heart of many critical technologies that span telecommunications,\cite{wang04.01} optical data storage,\cite{parte89.01} three-dimensional nano-photolithography,\cite{cumps99.01,kawat01.01} and making new materials\cite{karot04.01} for novel cancer therapies.\cite{roy03.01}  Because the strength of interaction, as quantified by the nonlinear-optical susceptibility, governs the efficiency of an application - and indeed whether or not a particular application is practical, making materials with ever larger nonlinear susceptibility has been the central focus of research in organic nonlinear optics.  Is there a fundamental limit to the susceptibility?  Our work shows that nature imposes such an upper bound on all optical phenomena.  Prior work determined the fundamental limit of only the {\em off-resonant} susceptibility.\cite{kuzyk00.02,kuzyk01.01,kuzyk00.01,kuzyk03.01,kuzyk03.02,kuzyk04.02}  In the present studies, we calculate the most general case: the fundamental limits of the dispersion of the real and imaginary parts of all electronic nonlinear-optical susceptibilities, which can be used to study or apply any nonlinear optical phenomena at any wavelength.  Our work provides an understanding of the constraints imposed by nature on our ability to control and use light, and lays the foundation for developing better materials and novel applications.

We focus on the second-order nonlinear-optical susceptibility of a molecule, often called the hyperpolarizability $\beta$, as an example of the process we use for finding the fundamental limit of any nonlinear susceptibility.  Furthermore, since our goal is to set an upper bound, we will only consider the largest tensor component, $\beta_{xxx}$.  The electronic nonlinear-optical susceptibilities are calculated using perturbation theory in the dipole approximation, which yields an expression that incudes the excited state properties of all the quantum states of the system.  This theoretical result is called a sum-over-states (SOS) expression, and for $\beta$ is given by:\cite{orr71.01}
\begin{eqnarray}\label{beta}
\beta_{xxx} (\omega_1, \omega_2) & = & -   \frac {e^3} {2} P_{\omega_{\alpha}, \omega_{\beta}} \left[ {\sum_{n}^{\infty}} ' \frac {\left| x_{0n} \right|^2 \Delta x_{n0}} {D_{nn}^{-1}(\omega_{\alpha},\omega_{\beta})} \right. \\ \nonumber
& + & \left.{\sum_{n }^{\infty}} ' {\sum_{m \neq n }^{\infty}} ' \frac {x_{0n} x_{nm} x_{m0}} {D_{nm}^{-1}(\omega_{\alpha},\omega_{\beta}) }\right],
\end{eqnarray}
where $-e$ is the electron charge, $x_{nm}$ the $n,m$ matrix element of the position operator, $\Delta x_{n0} = x_{nn} - x_{00}$ is the difference in the expectation value of the electron position between state $n$ and the ground state, $D_{nm}^{-1}(\omega_{\alpha},\omega_{\beta})$ gives the dispersion of $\beta$ (defined later) and $\hbar \omega_{\nu}$ are the photon frequencies.  The primes indicate that the ground state is excluded from the sum and the permutation operator $P_{\omega_{\alpha}, \omega_{\beta}}$ directs us to sum over all six frequency permutations.  Since the dipole moment of the molecule is proportional to the position ($p_x = -e x$), we loosely call $x_{nm}$ the transition moment and $x_{nn}$ the excited state dipole moment.   The first and second terms in Equation \ref{beta} are called the dipole and the octupolar terms.

Equation \ref{beta} is a function of an infinite number of material parameters, $x_{nm}$ and $E_{n0}$, so the maximum value of $\beta$ for each possible pair of photon energies requires an optimal set of transition moments and energies.  The sum rules, which are directly derivable from the Schr\"{o}dinger Equation, are relationships between the transition moments and energies.  In the present work, we apply the sum rules to put the SOS expression of the nonlinear optical susceptibilities into a form that can be maximized to calculate the ultimate nonlinear-optical susceptibility.

\section{Theory}

The generalized Thomas-Kuhn sum rules, derived directly from the Schr\"{o}dinger Equation without any approximations, relate the matrix elements and energies to each other according to:\cite{kuzyk01.01}
\begin{equation}
\sum_{n=0}^{\infty} \left( E_n - \frac {1} {2} \left( E_m + E_p \right) \right) x_{mn} x_{np} = \frac {\hbar^2 N} {2m} \delta_{m,p},
\label{sumrule}
\end{equation}
where $m$ is the mass of the electron, and $N$ the number of electrons.  The sum, indexed by $n$, is over all states of the system.  Equation \ref{sumrule} represents an infinite number of equations, one for each value of $m$ and $p$.  As such, we refer to a particular equation using the notation $(m,p)$.

Defining $E_{ij} = E_i - E_j$, we eliminate the dipole term using Equation $(m,p)$ with $m \neq p$:
\begin{equation}
\sum_{n=0}^{\infty} \left( E_{nm} + E_{np} \right) x_{mn} x_{np} =0 .
\label{sumrule m<>p}
\end{equation}
Equation \ref{sumrule m<>p} can be rewritten by explicitly expressing the $n=m$ and $n=p$ terms, setting $p=0$, and multiplying both sides by $x_{0m}$:\cite{kuzyk05.01a}
\begin{equation}
\Delta x_{m0} \left| x_{0m} \right|^2 = - {\sum_{n \neq m}^{\infty}} ' \frac {E_{nm} + E_{n0}} {E_{m0}}  x_{0m} x_{mn} x_{n0} .
\label{sumrule m<>p m,p removed dipole}
\end{equation}
Substituting Equation \ref{sumrule m<>p m,p removed dipole} into Equation \ref{beta}, we get,
\begin{eqnarray}\label{beta contracted}
\beta_{xxx} (\omega_1, \omega_2) & = & - \frac { e^3} {2} P_{\omega_{1}, \omega_{2}} \\ \nonumber & \times &
{\sum_{m}^{\infty}} ' {\sum_{n  \neq m}^{\infty}} ' \frac {x_{0m} x_{mn} x_{n0}} {D_{nm}^{-1} (\omega_1 , \omega_2)} \\ \nonumber
& \times & \left[ 1 - \frac {D_{nm}^{-1} (\omega_1 , \omega_2)} {D_{nn}^{-1} (\omega_1 , \omega_2)} \left( 2 \frac {E_{m0}} {E_{n0}} - 1 \right) \right] ,
\end{eqnarray}
where,
\begin{eqnarray}\label{beta denominators}
&& P_{\omega_{1}, \omega_{2}} D_{nm} (\omega_1 , \omega_2)  =  \frac {1} {2 \hbar^2} \left[ \frac {1} { \left(\omega_{ng} - \omega_1 - \omega_2 \right) \left(\omega_{mg} - \omega_1 \right)} \right. \\ \nonumber
& + &  \frac {1} {\left(\omega_{ng}^* + \omega_2 \right) \left(\omega_{mg} - \omega_1 \right)}  \\ \nonumber
& + &  \frac {1} {\left(\omega_{ng}^* + \omega_2 \right) \left(\omega_{mg}^* + \omega_1 + \omega_2 \right)} \\ \nonumber & + & \left. \omega_1 \leftrightarrow \omega_2 \hspace{1em} \mbox{for the three previous terms} \right],
\end{eqnarray}
and where $\omega_{mg} = \omega_{mg}^0 - i \Gamma_m /2$.  $\hbar \omega_{mg}^0$ is the energy difference between states $E_m$ and the ground state and $\Gamma_m$ is the damping width.  Equation \ref{beta contracted} is called the dipole-free expression or the reduced hyperpolarizability,\cite{kuzyk05.01a} where the second term in brackets implicitly accounts for the dispersion of all dipolar terms.

We use the ansatz that at most two states contribute to a molecule's nonlinear susceptibility near its fundamental limit.\cite{kuzyk05.02a,kuzyk05.01} (We stress that a three-level SOS model does not demand that the sum rules are truncated if the energies of the higher-lying state are high enough, so no truncation pathologies result.  Details will be described in a future publication.) Using Equation \ref{beta contracted},
\begin{eqnarray}\label{beta contracted-threelevel}
\beta_{xxx} (\omega_1, \omega_2) & = & P_{\omega_{1}, \omega_{2}} \mu_{01} \mu_{12} \mu_{20}  \\ \nonumber & \times & \left[  \frac {1} {D_{12}^{-1} (\omega_1 , \omega_2)} -  \frac { \left( 2 \frac {E_{20}} {E_{10}} -1 \right) } {D_{11}^{-1} (\omega_1 , \omega_2)} \right. \\ \nonumber & + & \left. \frac {1} {D_{21}^{-1} (\omega_1 , \omega_2)} - \frac { \left( 2 \frac {E_{10}} {E_{20}} -1 \right) } {D_{22}^{-1} (\omega_1 , \omega_2)} \right] \\ \nonumber 
& \equiv & \mu_{01} \mu_{12} \mu_{20} D^{3L} (\omega_1, \omega_2) ,
\end{eqnarray}
where $D^{3L} (\omega_1, \omega_2)$ is the term in brackets.

To further simplify Equation \ref{beta contracted-threelevel} so that we can calculate the extrema, we again use the sum rules to express the matrix element product $\mu_{01} \mu_{12} \mu_{20}$ in terms of $\mu_{01}$.  This leads to the exact result (without truncating the sum rules) is given by
\begin{equation}
\left| \mu_{01} \mu_{12} \mu_{20} \right| \leq \frac {E} {\sqrt{1-E}} \mu_{10}\sqrt{\mu_{MAX}^4 - \mu_{10}^4}, \label{momentProduct}
\end{equation}
where $E=E_{10}/E_{20}$ and $\mu_{MAX}$ is the maximum allowed transition dipole moment to the first excited state, which is related to the number of electrons, $N$, and the energy to the first excited state, $E_{10}$, according to
\begin{equation}
\mu_{MAX}^2 = \frac {e^2 \hbar^2} {2mE_{10} } N .
\label{groundsumrule}
\end{equation}

Having exhausted all the sum rules that meet the three criteria for physical soundness,\cite{kuzyk05.01,kuzyk05.02a} we assume that the remaining energies and transition moments are independent, so $\beta$ (via Equation \ref{momentProduct}) is maximal when $\mu_{10} = \sqrt[-4]{3} \mu_{MAX}$.  Combining Equations \ref{beta contracted-threelevel}, \ref{momentProduct} and \ref{groundsumrule}, with $\mu_{10} = \sqrt[-4]{3} \mu_{MAX}$, we get
\begin{equation}
\beta_{xxx} (\omega_1, \omega_2) \leq \beta_{0}^{MAX} \cdot \frac {1} {6} \frac {E E_{10}^2} {\sqrt{1-E}} \cdot D^{3L} (\omega_1, \omega_2)
\label{beta contracted-Dispersion}
\end{equation}
where $\beta_{0}^{MAX}$ is the fundamental limit of the off-resonant hyperpolarizability:\cite{kuzyk00.01}
\begin{equation}
\beta_{0}^{MAX} = \sqrt[4]{3} \left( \frac {e \hbar} {\sqrt{m}} \right)^3 \frac {N^{3/2}} {E_{10}^{7/2}} .
\label{beta-max-off-res}
\end{equation}
The maximum value of $\beta_{xxx}$ for any second-order process at any wavelengths $\omega_1$ and $\omega_2$ is given by Equation \ref{beta contracted-Dispersion} - a function of $E$, $E_{10}$, $\Gamma_{n0}$, and $N$.

\section{Discussion}

\begin{figure}
\scalebox{1}{\includegraphics{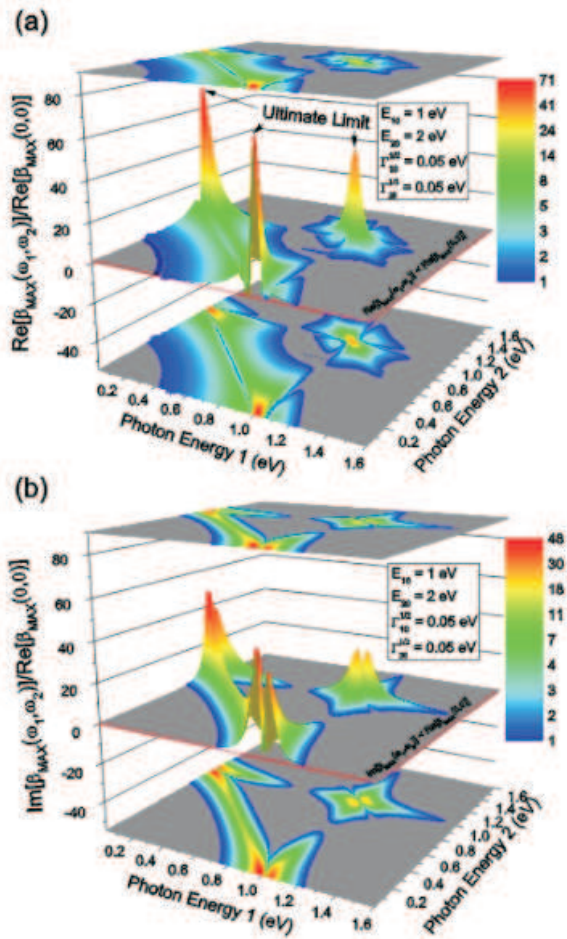}}
\caption{The dispersion of (a) the real part and (b) the imaginary part of fundamental limit of $\beta$ normalized to the off-resonant limit with $E_{10} = 1 eV$, $E_{20} = 2 eV$ and $\Gamma_{10}^{FWHM} = \Gamma_{20}^{FWHM} = 100 meV$ for a one-electron system.} \label{fig:ReBeta}
\end{figure}

All second-order nonlinear optical processes are bounded by Equation \ref{beta contracted-Dispersion} if the three-level ansatz is obeyed.  Figure \ref{fig:ReBeta} shows the real and imaginary parts of the maximum allowed hyperpolarizability as a function of the energy of the two incident photons, normalized to the maximum off-resonant value $\beta_0^{MAX}$.  Energies of the two dominant states used in these plots are $E_{10} = 1 eV$, $E_{20} = 2 eV$ and $\Gamma_{10}^{FWHM} = \Gamma_{20}^{FWHM} = 100 \, meV$ (a $100 \, meV$ width is a common approximation for organic molecules).

There are two dominant peaks in the fundamental limit of the real part of $\beta_{xxx}$.  The electrooptic hyperpolarizability, $\beta_{xxx} (0, \omega)$, peaks at just under 80 times the off-resonant fundamental limit while the second harmonic generation hyperpolarizability, $\beta_{xxx} (\omega, \omega)$, peaks at just under 50 times the off-resonant limit.  Using these excited state energies and damping factors, the ultimate hyperpolarizabilities for an $N$-electron system is
\begin{equation}
\beta^{ULTAMATE} = 91,000 \times 10^{-30} \frac {cm^5} {statcoul} \cdot N^{3/2} .
\label{beta-ultimate}
\end{equation}

For the {\em off-resonant fundamental limit} of $\beta(0,0)$, the second dominant excited state energy is required to be infinite.  Since the energy level spacing typically gets smaller for higher energy states, most molecules should fall far short of the limit.  Indeed, this could explain the factor of 30 gap between the best molecules and the off-resonant limit.\cite{Kuzyk03.05,Tripa04.01}

\begin{figure}
\scalebox{1}{\includegraphics{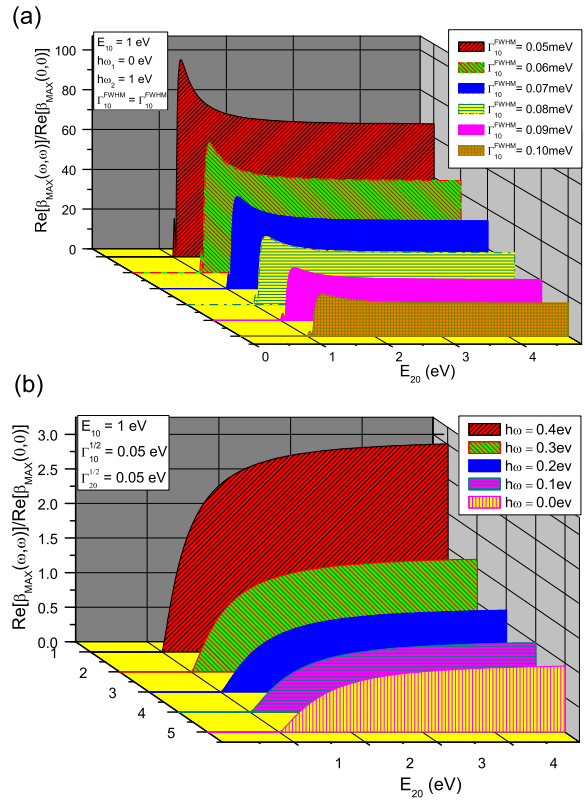}}
\caption{(a) The real part of fundamental limit of $\beta$ normalized to the off-resonant limit with $E_{10} = 1 eV$ as a function of $E_{20}$, for various values of $\Gamma_{10}^{FWHM} = \Gamma_{20}^{FWHM}$; and (b) The real part of the fundamental limit of the second harmonic $\beta$ normalized to the off-resonant limit with $E_{10} = 1 eV$ as a function of $E_{20}$, for various off-resonant pump wavelengths.} \label{fig:ReBetaVary}
\end{figure}

Figure \ref{fig:ReBetaVary}a shows a plot of the real part of the fundamental limit of the resonant electrooptic hyperpolarizability $\beta (0,\hbar \omega = E_{10})$ normalized to the off-resonant limit (with $E_{10} = 1 eV$) as a function of $E_{20}$, for various values of $\Gamma_{10}^{FWHM} = \Gamma_{20}^{FWHM}$.  As expected, the maximum resonant value increases as the width decreases.  However, the hyperpolarizability peaks when the second dominant excited state is close in energy to the first dominant excited state.  In the limit of zero width, $\beta$ peaks when the two dominant states are degenerate.  Such excited state energy spacing is common in organic molecules, so it may be possible to identify molecules with an ultra-large resonant response using this strategy.

The structure required of a molecule to be at the fundamental limit is thus different on and off resonance.  Our present calculations, in the zero-frequency limit of $\beta(\omega,\omega)$, are summarized by the series of plots in Figure \ref{fig:ReBetaVary}b, which show the fundamental limits for several photon energies between $0$ and $0.4 \, eV$ with a damping factor of $\Gamma_{n0} = 100meV $.  For $\hbar \omega = 0$, the fundamental limit clearly approaches the normalized value of unity, as previously calculated.  For each off-resonant photon energy, the fundamental limit increases as $\hbar \omega \rightarrow E_{10}/2$ and the second excited state energy approaches infinity.

The limits in the dispersion of the real and imaginary parts of the third-order susceptibility, $\gamma(\omega_1, \omega_2, \omega_2)$, can be calculated in a similar way using the three-level ansatz in dipole-free form.\cite{kuzyk05.01a} The sum rules can then be used to re-express the results in a form analogous to Equations \ref{beta contracted-Dispersion}.  A full discussion of third-order susceptibilities will be presented in a future paper.

It is interesting to apply our results to typical molecules.  For example, consider a molecule with 50 conjugated electrons whose first excited state energy is $E_{10} = 1.55 \, eV \, (800 \, nm)$ and $\Gamma = 100 \, meV$.  According to Equation \ref{beta-max-off-res},\cite{kuzyk00.01} the off-resonant value of the hyperpolarizability yields $\beta_0^{MAX} = 90,500 \times 10^{-30} \, esu$.  Using our new theoretical results, the resonant electrooptic hyperpolarizability is $\beta_{MAX}(0,\hbar \omega = E_{10}) = 7, 240,000 \times 10^{-30} esu$.  For a polar-aligned bulk material made of such molecules, at a concentration of $10^{21} \, cm^{-3}$, the fundamental limit of the off-resonant electrooptic coefficient is $r_{MAX}(0,0) = 3,000 \, pm/V$ and the resonant case yields $r_{MAX}(0,\hbar \omega = E_{10}) = 240,000 \, pm/V$.  Higher concentration systems, or ones with narrower widths, would have even higher values.  As such, our calculations show that there is room for considerable improvement over today's best materials.

Our theory is general and few approximations have been used in calculating the dispersion of the limits.  We start with a dipole-free form of the SOS expression for $\beta$ and apply the ansatz that the oscillator strength is dominated by two excited states when the susceptibility approaches the fundamental limit.  So in our calculation, {\em the three-level model is not an approximation}, but an exact model for a system at the limit.\cite{kuzyk05.02a,kuzyk05.01}  Furthermore, since this does not demand the sum rules to be truncated to three levels, no truncation pathologies result.  The fact that most real molecules are not well approximated by a three-level model may explain the factor-of-thirty gap between the fundamental limit and the best molecules.  The fact that a clipped harmonic oscillator yields a value of $\beta$ that is near the fundamental limit\cite{Tripa04.01} shows that the calculated limits are not large overestimates.  The set of all theoretical calculations and measurements are consistently below the fundamental limit, leading credence to the theory.

We only consider processes that are described by a sum-over-states expression which inherently assumes the electric dipole approximation (i.e. the wavelength of light is small compared with the size of the quantum system under study).  While it is often the case that higher-order terms or magnetic moments may be larger than lower-order terms, the {\em fundamental limits} of the higher-order terms should not be larger than the {\em fundamental limits} of lower-order terms.  Secondly, we only consider the largest diagonal components of $\beta$ and $\gamma$.  Short of unexpected new Physics, we would not expect that the maximum limit of an off-diagonal component of a susceptibility to exceed a diagonal one.  Note that our calculations include resonance, so Kleinman Symmetry is expected to be broken, leading to potentially large off-diagonal components of susceptibilities that are disallowed off resonance.  To treat such cases, one would need to use off-diagonal sum rules.

In conclusion, we have applied the sum rules to determine the fundamental limits of the dispersion of any complex susceptibility, thus being applicable to any electronic nonlinear-optical process.  We find that to attain the fundamental limit, off-resonant processes require a different molecular structure than on resonance, so one approach of molecular engineering does not suit all applications.  Our work can be applied to calculating and optimizing the figure of merit of materials for specific applications; provides a method for comparing measured values of $\beta$ and $\gamma$ without having to resort to determining $\beta_0$ and $\gamma_0$, which requires the use of unreliable dispersion models; and gives guidance for designing new materials and how their dispersion can be tuned for for a desired response.  Clearly, there is still room for substantial improvements in all nonlinear molecular susceptibilities, especially on resonance in systems with closely-spaced excited states, such as provided by octupolar molecules.\cite{Joffr92.01}

{\bf Acknowledgements: } I thank the National Science Foundation (ECS-0354736) and Wright Paterson Air Force Base for generously supporting this work.

%\bibliography{\bibs}

\clearpage

\end{document}